\documentclass[aps, onecolumn, showkeys, showpacs, superscriptaddress,groupedaddress, nofootinbib, floatfix, 11pt]{revtex4-2}  
\usepackage{graphicx}  
\usepackage{dcolumn}   
\usepackage{bm}        
\usepackage{amssymb}
\usepackage{amsmath}   
\usepackage{hyperref}
\hypersetup{colorlinks,linkcolor={blue},citecolor={blue},urlcolor={red}}
\usepackage{float}
\usepackage{xcolor}

\newcommand{\viz}{\textit{viz.~}}

\def\beq{\begin{equation}}
	\def\eeq{\end{equation}}
\def\bea{\begin{eqnarray}}
	\def\eea{\end{eqnarray}}

\begin{document}

\title{Precision Inflationary Predictions: Impact of Accurate End-of-Inflation Dynamics}

\author{Debottam Nandi$^{1}$}
\email{debottam.nandi@vit.ac.in}
\author{Simran Yadav$^2$}
\email{simranyadavkhola@gmail.com}
\author{Manjeet Kaur$^2$}
\email{mkaur1@physics.du.ac.in}
\affiliation{$^1$Department of Physics, School of Advanced Sciences, Vellore Institute of Technology (VIT) Chennai, Chennai 600127, India\\
	$^2$Department of Physics and Astrophysics, University of Delhi, Delhi 110007, India}


\begin{abstract}
The precision era of cosmology demands accurate theoretical predictions from inflationary models. In quantitative reheating analyses, inflationary observables depend sensitively on the number of e-folds between horizon exit and the end of inflation, $N_k$, whose determination relies on slow-roll approximations near the end of inflation. Since inflation ends when the first slow-roll parameter reaches unity, even modest inaccuracies in this approximation can shift the end of inflation and thereby alter $N_k$, leading to modifications in predicted observables---including those evaluated at leading-order. While such effects are implicit in standard treatments, their quantitative impact on observable constraints has not been systematically assessed. In this work, we first re-evaluate leading-order slow-roll predictions using an improved determination of $N_k$ within a simple quantitative reheating framework, and then incorporate higher-order slow-roll corrections consistently with the revised background evolution. Applying this framework to the Starobinsky model, we find that improved end-of-inflation dynamics alone can induce shifts of order $\Delta n_s \sim 10^{-3}$, while higher-order slow-roll corrections provide additional refinements at the $\sim 4 \times 10^{-4}$ level. The cumulative effect yields a maximum shift of $\Delta n_s \sim 1.2 \times 10^{-3}$ within the allowed reheating range. To our knowledge, this is the first systematic decomposition of end-of-inflation corrections and their	individual contributions to $n_s$ in the Starobinsky model, with 
implications for model discrimination in next-generation CMB surveys. These results demonstrate that an accurate determination of the end of inflation is essential for precision tests of inflationary models.
\end{abstract}

\maketitle


\section{Introduction}\label{sec:Introduction}
 
The inflationary paradigm \cite{Starobinsky:1979ty, Starobinsky:1980te, Guth:1981, Sato:1981, Mukhanov:1981xt, LINDE1982389, HAWKING1982295, Starobinsky:1982ee, Guth:1982, Sasaki1986, Albrecht-Steinhardt:1982, Linde:1983gd, VILENKIN1983527, Bardeen:1983, 1990-Kolb.Turner-Book, Mukhanov:1990me, Liddle:1994dx, Lidsey:1995np, Copeland:1997et, Martin:2000xs, 2005hep.L, Lyth:2007qh, Sriramkumar:2009kg, Baumann:2009ds, Martin:2013tda, Linde:2014nna, Martin:2015dha, Ade:2015lrj, Ade:2015ava, Clesse:2015yka, Odintsov:2023weg} provides a compelling description of the early universe, resolving the horizon and flatness problems while generating primordial perturbations in excellent agreement with observations of the cosmic microwave background and large-scale structure. In particular, slow-roll inflationary models predict an almost scale-invariant spectrum of fluctuations consistent with current data \cite{Planck:2018jri, Planck:2018vyg, BICEP:2021xfz, Galloni:2022mok}. As cosmological observations steadily improve in precision, however, the reliability of inflationary predictions increasingly depends on the accuracy of the underlying theoretical framework. In this precision-driven context, effects that are traditionally regarded as subleading can become observationally relevant, motivating a reassessment of standard approximation schemes.

Connecting inflationary predictions to observations requires accounting for the post-inflationary evolution of the universe, in particular the reheating epoch that bridges inflation and the radiation-dominated era \cite{Albrecht:1982mp, Abbott:1982hn, Traschen:1990sw, Kofman:1994rk, Shtanov:1994ce, Kofman:1997yn, Bassett:2005xm,   Allahverdi:2010xz, Martin:2010kz,  Mielczarek:2010ag,  Amin:2014eta, Dai:2014jja, Martin:2014nya, Domcke:2015iaa, Cook:2015vqa, Maity:2016uyn, Lozanov:2016hid, Kabir:2016kdh, Maity:2018qhi,  Maity:2019ltu, ElBourakadi:2021blc, Odintsov:2023lbb, German:2023yer}. In quantitative analyses \cite{Munoz:2014eqa, Maity:2018qhi, Martin:2006rs, Martin:2010kz, Adshead:2010mc, Mielczarek:2010ag, Dai:2014jja, Domcke:2015iaa, Gong:2015qha, Lozanov:2016hid, Nandi:2019xve, Liddle:2003as}, reheating is commonly characterized by an effective equation-of-state parameter and the duration of the reheating phase, which together determine the number of e-folds between horizon exit of observable modes and the end of inflation. Since inflationary observables, such as the scalar spectral index $n_s$ and the tensor-to-scalar ratio $r$, are highly sensitive to this duration, uncertainties in reheating directly propagate into predicted cosmological parameters.

Despite the sensitivity of inflationary predictions to post-inflationary evolution, it is commonly assumed that leading-order slow-roll approximations provide sufficiently accurate estimates of the inflationary background dynamics, including the determination of the end of inflation. In standard treatments, the end of inflation is inferred from slow-roll conditions, although inflation strictly terminates when the first slow-roll parameter reaches unity. Even modest inaccuracies in this approximation shift the end of inflation and therefore modify the duration of inflation. Since the number of e-folds $N_k$ between horizon exit and the end of inflation depends directly on this endpoint, a careful and consistent determination of the end of inflation becomes essential for reliable theoretical predictions.

The central role of $N_k$ becomes particularly evident in quantitative reheating analyses, where it appears explicitly in the consistency relation that connects the comoving pivot scale to the post-inflationary expansion history. More precisely, $N_k$ depends on both the effective reheating equation-of-state parameter $w_{\rm re}$ and the reheating duration $N_{\rm re}$, subject to the physical constraint $N_{\rm re} \geq 0$, where $N_{\rm re} \ll 1$ corresponds to instantaneous reheating. Through this relation,  $N_k$ determines the value of the inflaton field at horizon exit and hence the slow-roll parameters that enter predictions for observables such as the scalar spectral index $n_s$ and the tensor-to-scalar ratio $r$. Even within leading-order slow-roll expressions, small shifts in $N_k$ directly translate into corrections to $n_s$ and $r$ at a level comparable to current \cite{Planck:2018jri, Planck:2018vyg, BICEP:2021xfz, Galloni:2022mok} and forthcoming observations such as PRISM \cite{Andre:2013afa}, EUCLID \cite{Amendola:2012ys}, cosmic 21-cm surveys \cite{Mao:2008ug}, and CORE \cite{Finelli:2016cyd} sensitivities. As cosmological measurements continue to improve in precision, achieving reliable theoretical control over the determination of $N_k$ becomes increasingly important. Moreover, inflationary predictions and their post-inflationary reheating constraints have been explored across a wide variety of theoretical frameworks, including warm inflation with Chaplygin gas backgrounds \cite{Jawad:2024pmz, Maqsood:2021zwz}, warm tachyon scalar field inflation~\cite{Rani:2024zan}, warm inflation in $f(\phi, T)$ modified gravity~\cite{Alruwaili:2025daq}, and attractor models in extended teleparallel gravity~\cite{Qummer:2021xou} --- all of which confirm that precise control of inflationary observables and their post-inflationary history is essential for consistency with CMB data. However, in most standard treatments, the effect of accurately determining the end of inflation is implicitly assumed to be negligible and has not been systematically quantified.

Motivated by this sensitivity, the primary objective of this work is to obtain a more reliable determination of the duration of inflation and the corresponding number of e-folds $N_k$ by improving the treatment of the inflationary background dynamics near the end of inflation. Since inflation terminates when the first slow-roll parameter reaches unity, even modest inaccuracies in the slow-roll approximation can shift the end of inflation and thereby alter $N_k$, leading to modifications in inflationary observables --- including those evaluated at leading-order. While recent studies \cite{Kaur:2023wos, Auclair:2024udj} have indicated that improved treatments of the background evolution can induce non-negligible corrections to the estimation of $N_k$, these effects have not been systematically quantified with explicit numerical implications for inflationary observables --- nor has the individual contribution of each correction been isolated and decomposed separately. In this work, we address this gap by carrying out a detailed analysis in two stages. First, we reassess leading-order slow-roll predictions using an improved determination of $N_k$ within a simple and controlled quantitative reheating framework, thereby isolating the impact of end-of-inflation corrections on baseline theoretical predictions. Second, we incorporate higher-order slow-roll corrections \cite{Huang:2006yt, Caprini:2002jy, Cline:2006db, Huang:2006um, Huang:2006hr, Peiris:2006ug, Easther:2006tv, Peiris:2006sj, Izawa:2003mc, Ashoorioon:2005ep, Ballesteros:2005eg, Li:2006te, Chen:2006wn, Vennin:2015vfa, Karam:2017zno} consistently with the revised background evolution to evaluate additional refinements. As a representative and observationally viable benchmark, we apply this framework to the Starobinsky model of inflation \cite{Starobinsky:1979ty, Starobinsky:1980te, Starobinsky:1982ee, Starobinsky:1983zz, Starobinsky:1987zz} and  demonstrate that corrections often regarded as negligible can lead to 
appreciable modifications of inflationary predictions --- specifically, a 
cumulative shift of $\Delta n_s \sim 1.2 \times 10^{-3}$ in the scalar 
spectral index, dominated by the numerical background treatment ($\sim 10^{-3}$), with subdominant contributions from higher-order slow-roll terms 
($\sim 4 \times 10^{-4}$) and the revised reheating onset ($\sim 10^{-5}$). 
To our knowledge, this is the first systematic decomposition of these 
corrections and their individual propagation into $n_s$ for the Starobinsky 
model, with direct implications for model discrimination in next-generation 
CMB surveys operating at sensitivity $\Delta n_s \sim 10^{-3}$.

This article is organized as follows. In Sec.~\ref{sec: gen-eq}, we introduce the inflationary model and the corresponding background dynamics, including the slow-roll conditions and the associated approximations. Sec.~\ref{sec: reheating} presents the quantitative reheating analysis that connects inflationary perturbations to observable constraints, where the Starobinsky model is examined as a representative example using leading-order slow-roll expressions. In Sec.~\ref{sec: improvement}, we implement corrections to the inflationary dynamics and assess their impact on theoretical predictions and reheating constraints. Our conclusions are summarized in Sec.~\ref{sec:conclu}.

Throughout this work, we employ natural units with $\hbar = c = k_B = 1$ and set the reduced Planck mass to $M_{\rm pl} \equiv (8\pi G)^{-1/2} = 1$. The metric signature is $(-,+,+,+)$, Greek indices are contracted with the spacetime metric $g_{\mu\nu}$, and Latin indices with the Kronecker delta $\delta_{ij}$. We denote partial and covariant derivatives by $\partial$ and $\nabla$, respectively, while overdots and primes represent derivatives with respect to cosmic and conformal time in the Friedmann--Lema\^{\i}tre--Robertson--Walker background.

\section{General equations and slow-roll conditions}\label{sec: gen-eq}

In this section, we present the general background equations governing single-field inflation and review the standard slow-roll framework that underlies conventional analytical predictions. This formulation will serve as the baseline against which we later assess corrections arising from a more accurate treatment of the end of inflation. For that, let us consider the simplest action with a single canonical scalar field $\phi$ minimally coupled to gravity with a potential $V(\phi)$, which can be written as
\begin{eqnarray}\label{eq: action}
    S=\frac{1}{2}\int {\rm d}^4 x \sqrt{-g} \left[R - g^{\mu \nu} \partial_{\mu} \phi \partial_{\nu} \phi -2 V(\phi)\right].
\end{eqnarray}
Here, $R$ is the Ricci scalar. The corresponding equations of motion, i.e., Einstein's equations and the equation of the scalar field, can be written as
	\begin{eqnarray}\label{eq:eins-eq}
		R_{\mu\nu}-\frac{1}{2}g_{\mu\nu} R &=& T_{\mu\nu (\phi)},\\
		\label{eq:cont-eq}
		\nabla_\mu T^{\mu\nu}_{(\phi)}&=&0,
	\end{eqnarray}
	where $T^\mu_{\,\nu{(\phi)}}$ is the stress-energy tensor corresponding to the $\phi$ field and can be written as
	\begin{equation}\label{eq:EM-tensor}
		T_{\mu\nu{(\phi)}}=  \partial_\mu\phi\ \partial_\nu \phi-g_{\mu\nu}\left(\frac{1}{2}\partial_\lambda\phi \ \partial^\lambda \phi+V(\phi)\right).
	\end{equation}
Using the flat FLRW line element describing the homogeneous and isotropic universe at large scales, i.e., 
\begin{eqnarray}\label{eq: FLRW-metric}
    {\rm d}s^2 = -{\rm d}t^2 +a^2(t) {\rm d}\mathbf{x}^2,
\end{eqnarray}
where, $a(t)$ is the scale factor, Eqs. \eqref{eq:eins-eq} and \eqref{eq:cont-eq} can be reduced to
\begin{eqnarray}
	\label{eq: energy}
	&&3 H^2 = \frac{1}{2} \dot {\phi}^2 + V(\phi),\\
	\label{eq:acc-eq}
	&&\dot{H} = -\frac{1}{2}\dot{\phi}^2, \\
	\label{eq: phi}
	&&\ddot{\phi} + 3 H \dot{\phi} + V_{,\phi} = 0.
\end{eqnarray}
where, $H \equiv \dot{a}/a$ is the Hubble parameter, and $A_{x} \equiv \partial A/ \partial {x}$. These equations can be re-defined in terms of the e-fold variable $N \equiv \int H {\rm d} t = \ln \left(a/a_0\right)$ as
\begin{eqnarray}
\label{eq: H-n}
    &&H^2 = \frac{V}{3 - \frac{1}{2} \phi_N^2}, \qquad \frac{H_N}{H}= - \frac{1}{2} \phi_N^2,\\
\label{eq: phi-n}
    &&\phi_{NN} + \left(3 - \frac{1}{2}\phi^{2}_{N}\right) \left(\phi_N + \frac{V_\phi}{V}\right) = 0,
\end{eqnarray}  
where, $A_{x x} = \partial^2 A/ \partial {x}^2.$ For a given potential $V(\phi)$, these equations fully determine the background dynamics of inflation. In conventional analyses, these equations are typically simplified under the slow-roll approximation, yielding analytical expressions for inflationary observables. In the following, we briefly review this standard framework, which will serve as the baseline for the improved treatment developed later in this work.

Once the background equations are solved, one can in principle solve the scalar $(\mathcal{R}_k)$ and tensor $(h_k)$ perturbations and evaluate the power spectra:

\begin{eqnarray}\label{eq: pert}
	\mathcal{P}_{\mathcal{R}} = \frac{k^3}{2 \pi^2} |\mathcal{R}_k|^2 \equiv A_{\mathcal{R}}\left(\frac{k}{k_*}\right)^{n_s -1},\quad \mathcal{P}_T =2 .   \frac{k^3}{2 \pi^2} |h_k|^2 \equiv A_T\left(\frac{k}{k_*}\right)^{n_T}
\end{eqnarray}
where, $k_*$ is the pivot scale, $A_{\mathcal{R}},~A_T$  are the scalar and tensor spectral amplitude,
$$n_s \equiv \frac{d\, {\rm ln} \mathcal{P}_{\mathcal{R}}}{d\, {\rm ln} k}\bigg|_{k=k_*}, ~~~~~~~~~n_T\equiv \frac{d\, {\rm ln} \mathcal{P}_{T}}{d\, {\rm ln} k}\bigg|_{k=k_*}$$
 are the scalar and tensor spectral indices and
 $$r \equiv \frac{A_T}{A_{\mathcal{R}}}$$
is defined as the tensor-to-scalar ratio. The current observations constrain $A_{\mathcal{R}} \simeq 2.101^{+0.031}_{-0.034} \times 10^{-9} (68\%\, \, \text{CL})$, $n_{_{\rm s}} = 0.9672 \pm 0.0059\,(68\%\, \, \text{CL})$ and $r <0.028\, (95 \%\, \, \text{CL})$ at $k_*=0.05\, \text{Mpc}{}^{-1}$ (PLANCK \cite{Planck:2018jri, Planck:2018vyg, Galloni:2022mok}, BICEP/Keck \cite{BICEP:2021xfz} ). These observations agree remarkably well with the predictions of a near-scale-invariant spectrum of density perturbations in the inflationary paradigm. In the next section, we will discuss in detail how, without explicitly evaluating the perturbations, one can obtain these observables using the background solutions in the case of slow-roll inflation.

\subsection*{Slow-roll inflation}\label{sec: slow-roll dynamics}
We adopt the Hubble-flow definition of slow-roll parameters. Let us first define the two slow-roll parameters $\epsilon_{1}$ and $\epsilon_{2}$ as
	\begin{eqnarray}
		\label{eq:slow_roll_params-def}
		\epsilon_{\rm{1}} \equiv - \frac{\dot{H}}{H^2} = \frac{1}{2} \phi_N^2,\qquad 
		\epsilon_{\rm{2}} \equiv \frac{\dot{\epsilon_{\rm{1}}}}{H \epsilon_{\rm{1}}} = 2 \frac{\phi_{NN}}{\phi_N}.
	\end{eqnarray}
While $\epsilon_1<1$ ensures that the universe is accelerating, the condition $\epsilon_2\ll1$ ensures the sufficient duration of the inflationary phase. Inflation ends when the first slow-roll parameter reaches unity, i.e., $\epsilon_{1}=1$, without invoking slow-roll approximations. In the case of slow-roll, both these parameters are extremely small, i.e.,
\begin{eqnarray}\label{eq: slow-roll condition}
		&& \epsilon_{\rm{1}}\ll1, \quad \epsilon_{\rm{2}}\ll 1.
	\end{eqnarray}
These two conditions are also referred to as the slow-roll conditions. Under these conditions, Eqs. \eqref{eq: H-n} and \eqref{eq: phi-n} become 
\begin{eqnarray}
\label{eq: slow-roll-equations}
	H^2 \simeq \frac{1}{3} V, \quad \phi_{N} \simeq - \frac{V_{\phi}}{V},
\end{eqnarray}
and, using the above equation, the slow-roll parameters can be re-expressed in terms of shape of potential as
\begin{eqnarray}
\label{eq: slow-roll-para-approx}
	\epsilon_1 \simeq \frac{1}{2}\left(\frac{V_\phi}{V}\right)^2, \quad \epsilon_2 \simeq 2 \left(\frac{V^2_\phi}{V^2} - \frac{V_{\phi \phi}}{V} \right).
\end{eqnarray}
Eqs. \eqref{eq: slow-roll-equations} are referred to as the slow-roll equations. By solving these simplified equations, one can obtain the approximate analytical solution of the Hubble parameter as well as the slow-roll parameters, and thus the dynamics of the universe in slow-roll regime.


In order to check for consistency with the observation, one can, in principle, obtain the observables corresponding to the perturbations associated with the pivot scale $k$ in terms of slow-roll parameters as  
\begin{eqnarray}
\label{eq:cmb}
	A_{\mathcal{R}} \simeq \frac{H^2}{8 \pi^2 \epsilon_1}, ~~~~\quad n_s \simeq 1 - 2 \epsilon_1 - \epsilon_2, ~~~~\quad r \simeq 16 \epsilon_1.
\end{eqnarray}
The above expressions are obtained using leading-order slow-roll approximation, assuming the higher-order contributions are heavily suppressed. These observables depend explicitly on the e-folding number $N_k$, defined as the interval between horizon exit of the pivot scale and the end of inflation. Since $N_k$ is determined by the precise end of inflation, any inaccuracy in the slow-roll estimation of the end-of-inflation propagates directly into these leading-order predictions. Typically, this duration is considered to be $N_k \sim 50 - 60$. However, as we shall demonstrate, even shifts of order unity in $N_k$ can translate into corrections in $n_s$ at the level of $10^{-3}$, which is comparable to present and projected observational precision. However, it is a model-dependent quantity and in order to obtain the exact duration $N_k$, one must consider the post-inflationary dynamics, i.e., the reheating epoch. A consistent determination of $N_k$ therefore requires incorporating the post-inflationary expansion history. In the next section, we implement this through the quantitative analysis of reheating.

\section{Quantitative analysis of reheating}
\label{sec: reheating}
As mentioned in the previous section, in the slow-roll regime, the slow-roll parameters are very small, i.e., $\epsilon_1,\ \epsilon_2 \ll 1$. However, as the field rolls down the potential, these slow-roll parameters increase, and the inflation ends as soon as this condition is violated, i.e., $\epsilon_1 = 1$.  Briefly, after the end of inflation, the field reaches the bottom and begins to oscillate around it and couples with the other (standard) particles. As a consequence, at this stage, the time average of kinetic energy is the same as the average of the potential energy, and the field decays. This epoch is known as the reheating phase.\footnote{It is worth noting that in warm inflationary scenarios~\cite{Berera:1995fj, Berera:1995ie, Berera:2008ar, BasteroGil:2012cm, Bartrum:2013fia, Jawad:2024pmz, Rani:2024zan, Maqsood:2021zwz, Alruwaili:2025daq, Qummer:2021xou}, the inflaton continuously transfers energy to a thermal radiation bath \textit{during} inflation itself, so the distinct post-inflationary reheating phase analyzed here does not apply; the present work focuses exclusively on standard cold inflation followed by a separate reheating era, characterized by an effective equation-of-state parameter $w_{\rm re}$ and duration $N_{\rm re}$. A study of how end-of-inflation dynamics interplay with dissipation in warm scenarios is left to future work.}

This epoch involves intricate microphysical processes and is largely model-dependent. As a result, any analytical extension of its dynamics carries substantial uncertainties, making it extremely challenging to describe the epoch through a single, unified analytical framework. To circumvent this difficulty and achieve a model-independent treatment, a simplified quantitative approach has been developed in which the equation-of-state parameter during the entire reheating epoch, i.e., $w_{\rm re}$ is assumed to remain constant. Although this approximation does not capture the full microphysical complexity of reheating, it provides a useful average macrophysical description that successfully connects inflationary predictions with observations and thereby enables broad, indirect constraints on the reheating scenario. In this case, the energy density during reheating effectively behaves as

\begin{eqnarray}
	\rho \propto a^{- 3  (1 + w_{\rm re})}.
\end{eqnarray}
This relation follows from energy conservation $\dot{\rho} + 3 H (1 + w) \rho = 0$. We also define the duration of reheating $N_{\rm re}$ as
\begin{eqnarray}
	N_{\rm re} \equiv {\rm ln} \left(\frac{a_{\rm re}}{a_{\rm end}}\right)
\end{eqnarray}
where $a_{\rm end},\ a_{\rm re}$ are the scale factor solutions at the end of the inflation and reheating epoch, respectively. Physical consistency requires $N_{\rm re} \geq 0,$ where $N_{\rm re} \ll 1$ corresponds to instantaneous reheating.
In order to calculate the duration $N_{\rm re}$, let us compare the mode of interest with the present scale as
\begin{eqnarray}\label{eq: mode-expansion}
\frac{k}{a_0 H_0}=\frac{a_k H_k}{a_0 H_0}=\frac{a_k}{a_{\rm end}}\frac{a_{\rm end}}{a_{\rm {re}}}\frac{a_{\rm {re}}}{a_{\rm {eq}}}\frac{a_{\rm {eq}}}{a_{0}}\frac{H_{k}}{H_{\rm {eq}}}\frac{H_{\rm {eq}}}{H_{0}},
\end{eqnarray}
where $k$ is the comoving scale, and (${\rm eq}$) and ($0$) denote the quantities at the matter-radiation equality and the present epoch, respectively.  Then the above equation leads to:
\begin{eqnarray}\label{eq: k-scale}
{\rm ln}\left(
\frac{k}{a_0 H_0}\right)=-N_k-N_{\rm re}-N_{\rm {RD}}+{\rm ln}\left(\frac{a_{\rm {eq}}H_{\rm {eq}}}{a_{0}H_0}\right)+{\rm ln}\left(\frac{H_{k}}{H_{\rm {eq}}}\right),
\end{eqnarray}
where $N_k \equiv {\rm ln} \left(\frac{a_{\rm end}}{a_{k}}\right), \, N_{\rm RD} \equiv {\rm ln} \left(\frac{a_{\rm eq}}{a_{\rm re}}\right).$ At the end of reheating, the energy density is
\begin{eqnarray}\label{eq: re-energy}
\rho_{\rm {re}}=\frac{\pi^2}{30}g_{\rm {re}}\,T_{\rm {re}}^4,
\end{eqnarray}
where, $g_{\rm {re}}$ is the effective number of relativistic species upon thermalization. Radiation drives the ensuing expansion, with non-relativistic matter and dark energy making recent additions. Assuming a negligible entropy shift following $T_{\rm {re}}$, the current CMB and neutrino background both retain the reheating entropy, which leads to the relation, 
\begin{eqnarray}\label{eq: temp-re}
g_{\rm {s,re}}\,T_{\rm {re}}^3=\left(\frac{a_0}{a_{\rm {re}}}\right)^3 \left(2 T_0^3+6 \times \frac{7}{8}T_{\nu 0}^3\right),
\end{eqnarray}
where the present (CMB) temperature, $T_0 \simeq 2.73 $~K, the neutrino temperature, $T_{\nu 0} = (4/11)^{1/3} \, T_0$, and $g_{\rm s,re}$ is the effective number of species for entropy at reheating. Therefore, $T_{\rm {re}}$ can be written as the present temperature $T_0$ as 
\begin{eqnarray}
\label{eq: reh temp}
\frac{T_{\rm re}}{T_0}=\frac{a_0}{a_{\rm {eq}}}\frac{a_{\rm {eq}}}{a_{\rm {re}}}\left(\frac{43}{11 g_{\rm {s,re}}}\right)^{1/3},
\end{eqnarray}
and the Eq. \eqref{eq: k-scale} can be re-written as
\begin{eqnarray}
\label{eq: re_duration_final}
 N_{\rm re} = \frac{4}{1 - 3 w_{\rm re}} \left[-N_k - \frac{1}{4} \ln \left(\frac{30}{\pi^2 g_{\rm {re}}}\right) - \frac{1}{3} \ln \left(\frac{11 g_{\rm {s,re}}}{43}\right) - \ln\left(\frac{k}{a_0 T_0}\right) - \frac{1}{4} \ln \left(\frac{\rho_{\rm {end}}}{H_k^4}\right) \right]. \nonumber\\
 \end{eqnarray}
and $T_{\rm re}$ can be reduced to the following form:
\begin{eqnarray}
\label{eq: re temp end}
T_{\rm {re}} = \frac{a_0 T_0}{k}\left(\frac{43}{11 g_{\rm {s,re}}}\right)^{1/3} H_k e^{-N_{\rm k} - N_{\rm {re}}}.
\end{eqnarray}
where $H_k$ can be written as 
\begin{eqnarray}\label{eq: hubble parameter}
	H_k = \pi \sqrt{\frac{r A_{\mathcal{R}}}{2}}.
\end{eqnarray}
Please note that the Hubble parameter $H_k$ is a function of duration $N_k,$ and therefore, $N_{\rm re}$ is eventually a function of $N_k$. As spectral index $n_s$ (or any time-dependent parameters like the field $\phi$ or tensor-to-scalar ratio $r$) depends on $N_k$, considering $\{ g_{\rm re}, g_{\rm s,re}\} \sim
100 $, one can plot $n_s$ vs. $N_{\rm re}$ (or $T_{\rm re}$ from the above expression) and obtain a strict constraint on $N_{\rm re}$ \cite{Munoz:2014eqa, Martin:2013tda, Dai:2014jja, Kabir:2016kdh, Koh:2018qcy, Maity:2019ltu, deHaro:2023xcc} as  $n_s$ is strictly bounded. This is referred to as the quantitative analysis of reheating. In the next section, we will elaborately discuss this analysis by implementing it on the Starobinsky model of inflation.

\subsection{Starobinsky inflation}\label{sec: staro}
One of the most successful models of slow-roll inflation is the Starobinsky model \cite{Starobinsky:1979ty, Starobinsky:1980te, Starobinsky:1982ee, Starobinsky:1983zz,  Starobinsky:1987zz} where the potential in the Einstein frame is given by: 
\begin{eqnarray}
\label{eq: staro pot}
    V(\phi) = \frac{3}{4} M^2 \left(1- e^{-\sqrt{2/3} \phi}\right)^2,
\end{eqnarray}
where $M$ is the mass of the scalar field $\phi$,  whose value is constrained by the normalization of the scalar power spectrum to the CMB amplitude $\mathcal{A}_{\mathcal{R}} \simeq 2.1 \times 10^{-9},$ yielding $M \approx 1.3 \times 10^{-5} M_{\rm pl}.$
\begin{figure}[ht]
    \centering
    \includegraphics[width=0.6\textwidth]{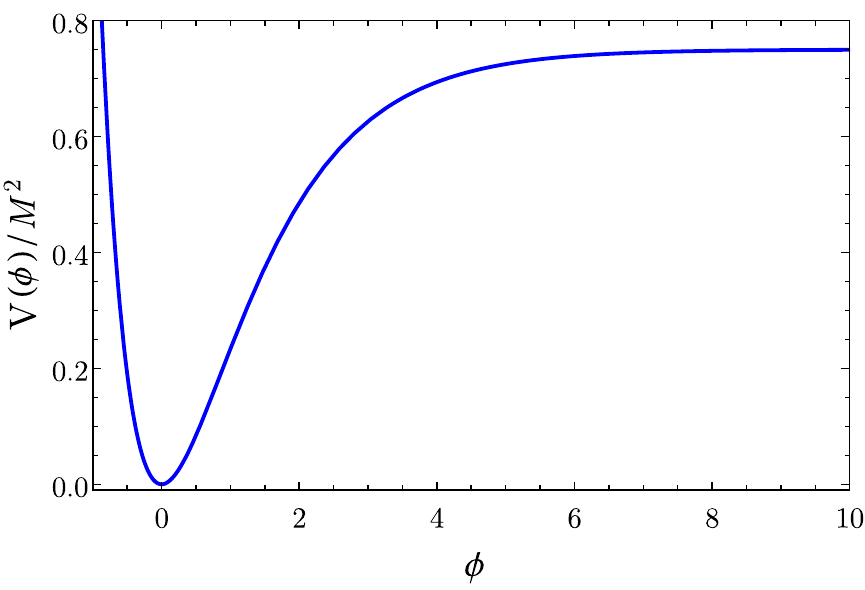} 
    \caption{The Starobinsky potential given in Eq. \eqref{eq: staro pot}.}
    \label{fig: starobinsky potential}
\end{figure}
Assuming the slow-roll conditions are satisfied, we can immediately obtain the two slow-roll equations (i.e., Eqs. \eqref{eq: slow-roll-equations}) as

\begin{eqnarray}\label{eq: staro dynamic eqn}
 H \simeq \frac{M}{2}\left(1- e^{-\sqrt{2/3} \phi}\right) , \quad  \phi_{N} \simeq {-2} \sqrt{\frac{2}{3}}\frac{e^{-\sqrt{2/3} \phi}}{\left(1-e^{-\sqrt{2/3} \phi}\right)},
\end{eqnarray}
and the slow-roll parameters can be expressed in terms of the field $\phi$ as (i.e., Eqs. \eqref{eq: slow-roll-para-approx}): 

\begin{eqnarray}\label{eq: staro slow-roll parameters}
 \epsilon_1 \simeq \frac{4 e^{-2\sqrt{2/3}\phi}}{3 \left(1 - e^{-\sqrt{2/3} \phi}\right)^2}, \qquad\quad \epsilon_2 \simeq \frac{8 e^{-\sqrt{2/3}\phi}}{3 \left(1 - e^{-\sqrt{2/3} \phi}\right)^2}.
\end{eqnarray}
With the help of Eq. \eqref{eq: staro dynamic eqn}, one can easily solve the scalar field solution in terms of the e-folding number $N_k,$ and the solutions of the slow-roll parameters at the leading-order, for $N_k \gg 1,$ can be obtained as:

\begin{eqnarray}
\label{eq: starobinsky slow-roll parameters}
\epsilon_1 \simeq \frac{3}{4 N_{k}^2}, \qquad\quad \epsilon_2 \simeq \frac{2}{N_{k}},
\end{eqnarray}
and, using the leading-order slow-roll approximation, i.e., Eqs. \eqref{eq:cmb}, we can quickly evaluate the scalar spectral index and the tensor-to-scalar ratio in terms of the e-folding number as:
\begin{eqnarray}
\label{eq: starobinsky observable parameters}
    n_{s} \simeq 1 - \frac{2}{N_{k}}, \qquad\quad r \simeq \frac{12}{N_{k}^2}.
\end{eqnarray}
\begin{figure}
    \centering
\includegraphics[width=0.6\textwidth]{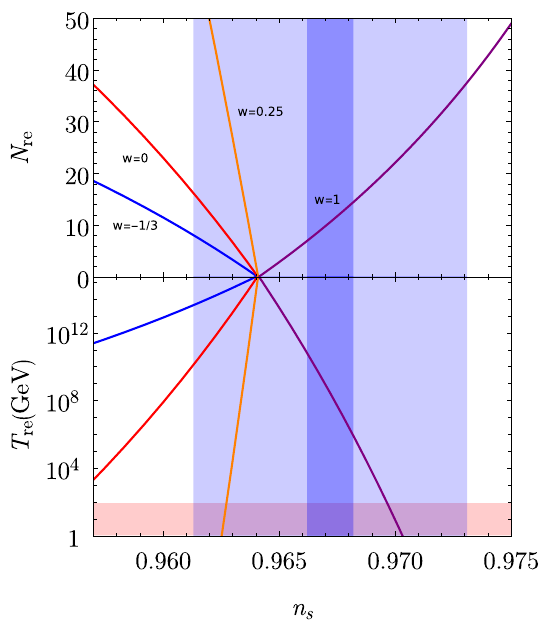}
    \caption{
    We plot the duration of reheating $N_{\rm re}$ and reheating temperature $T_{\rm re}$ given in Eqs. \eqref{eq: re_duration_final} and \eqref{eq: re temp end} as functions of the scalar spectral index $n_{s}$ given by Eq. \eqref{eq: starobinsky observable parameters} using leading-order slow-roll approximations for the case of Starobinsky inflation. Please note that different colors represent dynamics corresponding to different effective EoS parameter $w_{\rm re}$ as indicated in the figure. The blue-shaded region represents the $1\sigma$ constraint on the value of $n_s$ using ongoing observations \cite{Planck:2018jri, Planck:2018vyg, BICEP:2021xfz, Galloni:2022mok} with $n_{s} = 0.9672 \pm 0.0059$. The dark blue region shows the future projected bound on $n_s$  with a sensitivity of $10^{-3}$, assuming its central value remains unchanged. The temperature below the lighter red region is excluded due to the constraint from the electroweak scale, which is taken to be $100$ GeV.}
    \label{fig: starobinsky Nre_Tre}
\end{figure}
\begin{figure}
    \centering
\includegraphics[width=0.6\textwidth]{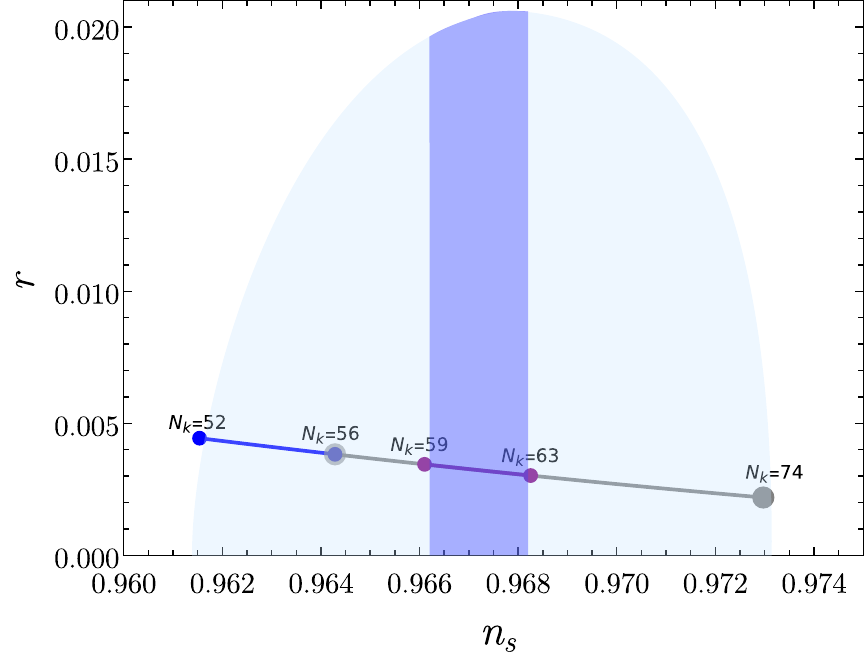}
    \caption{
    We plot the inflationary observables, tensor-to-scalar ratio ($r$) as a function of scalar spectral index ($n_s$) for the Starobinsky model using leading-order slow-roll approximations given by Eq. \eqref{eq: starobinsky observable parameters}. This evolution is presented considering the bound on duration $N_{ k}$ using the reheating regime. We consider different EoS parameters during reheating $w_{\rm re}$ and observe the constraint on the duration $N_{\rm k}$. Correspondingly, in the figure, blue line corresponds to the bound for $w_{\rm re}<1/3$, the gray line for $w_{\rm re}>1/3$ and the purple line for the future observational bound of $n_{\rm s}$. The blue-shaded region represents the $1\sigma$ constraint on the value of $n_{\rm s}$ using ongoing observations \cite{Planck:2018jri, Planck:2018vyg, BICEP:2021xfz, Galloni:2022mok} with $n_{\rm s} = 0.9672 \pm 0.0059$. The dark blue region shows the future projected bound on $n_{\rm s}$ with a sensitivity of $10^{-3}$, assuming its central value remains unchanged.}
    \label{fig: starobinsky ns_r}
\end{figure}
Using the above equation along with Eqs. \eqref{eq: re_duration_final} and \eqref{eq: re temp end}, one can obtain parametric dependence of $N_{\rm re}$ and $T_{\rm re}$ on $n_s$. In the case of the Starobinsky model, this is shown in Fig. \ref{fig: starobinsky Nre_Tre}. Keep in mind that, $N_{\rm re} \geq 0$, with $N_{\rm re} \ll 1$ is referred to as instantaneous reheating \cite{Felder:1998vq, deHaro:2023xcc}. It immediately leads to the constraint on the value of $n_s.$ For example, considering that during reheating, the evolution of the universe effectively behaves as matter-expansion, i.e., $w_{\rm re} = 0$. Thus, in the $ 1\sigma$ domain, we get the bound on $N_k$ as 
\begin{eqnarray}
    52 \leq N_k \leq 56
    \label{eq: staro Nk 1s bound}
\end{eqnarray}
for $w_{\rm re} = 0,$ and correspondingly, the bound on $n_s$ becomes:
\begin{eqnarray}
    0.961 \leq n_s \leq 0.964.
    \label{eq: staro ns 1s bound}
\end{eqnarray}
These bounds follow from imposing the observational constraint on $n_s$ together with the physical requirement $N_{\rm re} \geq 0$. In fact, as can be seen, for $w_{\rm re} < 1/3,$ it is obvious that there is an upper bound on $N_k$, which implies a bound on $n_s$ as well, i.e., $N_k \leq 56$ and $n_s < 0.964.$ It also immediately puts a tight constraint on $N_{\rm re}$ and $T_{\rm re}$ (for $w_{\rm re} = 0$) as
\begin{eqnarray}
    N_{\rm re} \leq 16, ~~~\quad T_{\rm re} \geq 10^{10} \,\text{GeV}.
    \label{eq: staro w<1/3 bound}
\end{eqnarray}
However, during reheating, if we assume $w_{\rm re} > 1/3,$ these constraints flip and in the $1\sigma$ domain, there arises a new bound with a lower value of $N_k$ as well as in $n_s$ as
\begin{eqnarray}
    56 \leq N_k \leq 74, ~~~~\quad 0.964\leq n_s \leq 0.973.
    \label{eq: staro w>1/3 Nk, ns bound}
\end{eqnarray}
Consequently, the bound on reheating parameters becomes (for $w_{\rm  re}=1$)
\begin{eqnarray}
    N_{\rm re} \leq 37, ~~~~\quad T_{\rm re} \geq 10^{-9} \,\text{GeV}.
    \label{eq: staro w>1/3 bound}
\end{eqnarray}
Note that reheating temperatures below the BBN scale $(\sim 10~{\rm MeV})$ are observationally excluded; therefore, the region corresponding to $T_{\rm re} \leq 10^{-2}$ GeV is not physically viable and must be discarded. In addition, considerations related to the electroweak scale $(\sim 100~{\rm GeV})$ may motivate a higher reheating temperature in specific baryogenesis scenarios; however, this requirement is model-dependent and does not provide a constraint as robust as the BBN bound. Taking these bounds into account, we plot $r$ as a function of $n_s$ in Fig. \ref{fig: starobinsky ns_r}. Further, instead of looking in $1\sigma$ limit, if we consider the observational constraints proposed by the forthcoming experiments such as PRISM \cite{Andre:2013afa}, EUCLID \cite{Amendola:2012ys}, cosmic 21-cm surveys \cite{Mao:2008ug}, and CORE \cite{Finelli:2016cyd} experiments, which offer a precision enhancement of $10^{-3}$ in $n_s$, the constraints on $N_k$, $N_{\rm re}$ and $T_{\rm re}$ can be substantially improved\footnote{The central value of $n_s$ is assumed to be $0.9672$ \cite{Galloni:2022mok}}. Then one can immediately refer from the Fig. \ref{fig: starobinsky Nre_Tre} and \ref{fig: starobinsky ns_r} that for $w_{\rm re} < 1/3,$ the scalar spectral index does not satisfy the constraints in $1\sigma$ level posed by future observations. Therefore, under these conditions, this model would be disfavored at the $1\sigma$ level if the central value remains unchanged. The model can survive if one assumes $w_{\rm re} > 1/3,$ and  the bound on $N_{\rm re}$ and $T_{\rm re}$ for the constraints posed by future observations becomes
\begin{eqnarray}
    7 \leq N_{\rm re} \leq 14, ~~~~~\quad 10^{9}\,\text{GeV} \geq T_{\rm re} \geq 10^{6}\,\text{GeV}.
    \label{eq: staro future obs bound}
\end{eqnarray}
Even in that case, if the future observations detect instantaneous reheating \cite{Felder:1998vq, deHaro:2023xcc}, i.e., $N_{\rm re} \ll 1,$ $n_s$ appears to be outside of the contour from the future observations. Therefore, unless $w_{\rm} > 1/3$ with the prolonged reheating era, the analytical approximation may lead to ruling out the Starobinsky inflation model. However, keeping aside the future observations, in the next section, we will see a significant improvement in these constraints by implementing meaningful corrections to the dynamics rather than the analytical approximations.

\section{Numerical improvements}\label{sec: improvement}

In the previous section, we discussed the CMB constraints on various inflationary models, which depend on the effective EoS parameter $w_{\rm re}$ as well as the duration of the reheating epoch $N_{\rm re}$. There are two issues associated with this analysis. First, around and at the end of inflation, the slow-roll approximation breaks down. Yet, in evaluating duration $N_k$ as well as energy density at end of inflation $\rho_{\rm end}$ in Eq. \eqref{eq: re_duration_final}, we rely upon slow-roll approximations. Second, in evaluating perturbed observables in Eqs. \eqref{eq:cmb}, we consider only the leading-order contributions and ignore higher-order contributions, assuming their impact to be negligible. However, as anticipated in Refs. \cite{Kaur:2023wos, Auclair:2024udj}, considering accurate dynamics rather than the slow-roll approximations may lead to notable enhancements in these predictions. Additionally, one can consider the onset of reheating not at the end of inflation but at the bottom of the potential, as will be discussed in the subsequent section. The implementation and, therefore, the resulting corrections due to these considerations can be categorized into three parts:
\begin{enumerate}
    \item Implementation of accurate inflationary dynamics through numerical methods instead of slow-roll approximations.
    \item Implementation of higher-order slow-roll approximations instead of leading-order slow-roll approximations.
    \item Implementation of onset of reheating as the bottom of the potential instead of the end of inflation.
\end{enumerate}
In the following sections, we will conduct a detailed study of the impact of each of these corrections on the observational constraints. We will also compare the cumulative effects of these corrections with the existing predictions presented in the previous section.

\subsection{Implementation of accurate inflationary dynamics through numerical methods}\label{sec:first_improv}

As mentioned earlier, this epoch of the slow-roll regime lasts as long as both of the slow-roll parameters are small, i.e., ($\epsilon_1\ll1,~\epsilon_2\ll1$). On the other hand, $\epsilon_1=1$ signals the end of inflation. It turns out that even before the end of inflation, these slow-roll approximations break down. Despite this, we still rely on (leading-order) slow-roll approximations even at the end of inflation to determine the analytical solution of the background variables, assuming the contribution due to accurate dynamics in evaluating perturbed observables is negligible. While slow-roll remains an excellent approximation during most of the inflationary phase, its validity deteriorates near the end of inflation where $\epsilon_{1} \rightarrow1$, and the higher-order kinetic contributions become non-negligible. However, it is crucial, as in this section, we shall learn the significance of obtaining the accurate values of not only the background parameters but also the end of inflation. To illustrate this, let's now discuss this in detail with an example: the Starobinsky inflation.

\begin{figure}[t]
    \centering
      \includegraphics[width=0.6\textwidth]{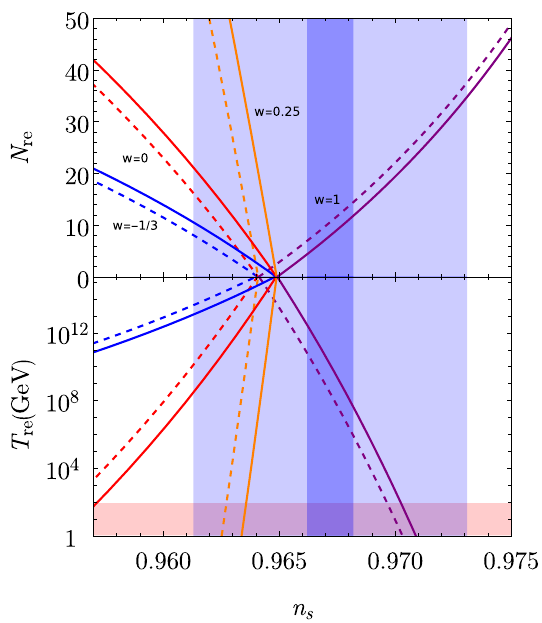}
  \caption{    We plot the duration of reheating $N_{\rm re}$ and reheating temperature $T_{\rm re}$ given by Eqs. \eqref{eq: re_duration_final} and \eqref{eq: re temp end} as functions of the scalar spectral index $n_{s}$ parametrically for both analytical and numerical solution, given by Eqs. \eqref{eq: starobinsky observable parameters} and \eqref{eq:cmb}, respectively. The solid lines are for the numerical solution, and the dashed lines are for the analytically approximated solution. Please note that different colors represent dynamics corresponding to different effective equations of state parameter $w_{\rm re}$ as indicated in the figure. The blue-shaded region represents the $1\sigma$ constraint on the value of $n_s$ using ongoing observations \cite{Planck:2018jri, Planck:2018vyg, BICEP:2021xfz, Galloni:2022mok} with $n_{s} = 0.9672 \pm 0.0059$. The dark blue region shows the future projected bound on $n_s$  with a sensitivity of $10^{-3}$, assuming its central value remains unchanged. The temperature below the lighter red region is excluded due to the constraint from the electroweak scale, which is taken to be $100$ GeV.
  }
    \label{fig: staro Nre_Tre ana num leading order}
\end{figure}

Using leading-order slow-roll approximations and solving the slow-roll equations (\viz Eqs. \eqref{eq: slow-roll-equations}):

$$H^2 \simeq \frac{V}{3}, \quad \phi_N \simeq - \frac{V_{\phi}}{V},$$
with the initial condition $\phi_i = 5.8,$ we obtain the following results:

$$N_{\rm end} \simeq 80.86,\quad \phi_{\rm end} \simeq 0.94, \quad H_{\rm end} \simeq 0.267\, M, \quad \rho_{\rm end} \simeq 0.215 \, M^2,$$
along with the slow-roll solutions given in Eqs. \eqref{eq: starobinsky slow-roll parameters}. In both the analytical and numerical cases, we impose the identical initial condition $\phi_i = 5.8$ to ensure a consistent comparison. Here, $N_{\rm end}$ is the duration of the field rolling from the initial value $\phi_i$ to $\phi_{\rm end},$ i.e., the duration of inflation. These values are essential for the analysis in Eqs. \eqref{eq: re_duration_final} and \eqref{eq: re temp end} which has been shown in the previous section with Fig. \ref{fig: starobinsky Nre_Tre}.

To improve the analysis, in this section, we now consider, rather than the slow-roll, the full inflationary equations (\viz Eqs. \eqref{eq: H-n} and \eqref{eq: phi-n}): 

$$H = \sqrt{\frac{V}{3 - \frac{1}{2}\phi_N^2}}, \quad \phi_{NN} + \left(3 - \frac{1}{2} \phi_N^2\right) \left(\phi_N + \frac{V_{\phi}}{V}\right) =0.$$

\noindent Solving the highly non-linear differential equation with identical initial condition $\phi_i = 5.8$  by using fourth-order Runge-Kutta (RK4) with adaptive step-size control, we now obtain the numerical solutions of the background variables such as the Hubble parameter and the slow-roll parameters along with:

$$N_{\rm end} \simeq 82.48, \quad \phi_{\rm end} \simeq 0.61, \quad H_{\rm end} = 0.241\,M, \quad \rho_{\rm end} \simeq 0.175\,M^2.$$

\noindent Thus, the numerical treatment shifts the end of inflation by $\Delta N_{\rm end} \simeq 1.6$, which directly propagates into the reheating analysis. These improvements directly impact the reheating analysis in two ways. The first is the direct influence of $\rho_{\rm end}$ in Eq. \eqref{eq: re_duration_final}. The second, and more crucial, impact is the significant improvement in $\Delta N_{\rm end} = \Delta N_k \sim 1.5,$ which in turn enhances the accuracy of the background variables. For example, in the case of leading-order approximations, $\{\epsilon_1, \epsilon_2\} \simeq  \{0.0003, 0.040\}$ for $N_k = 50,$  whereas, numerically we obtain these values as $\{\epsilon_1, \epsilon_2\} \simeq  \{0.00027, 0.038\},$ leading to a representative shift of $\Delta n_s \simeq 2 \times 10^{-3}$ around $N_k \simeq 50,$ as also discussed in Ref. \cite{Kaur:2023wos}~\footnote{Please note that, for a multi-parameter model, numerical evaluation for the entire range of the parameters can be challenging. In that case, one can rely upon a semi-analytical solution presented in Ref. \cite{Kaur:2023wos}, where analytically obtained solutions are nearly comparable to the numerical solutions.}.

\begin{figure}[h!]
    \centering
     \includegraphics[width=0.6\textwidth]{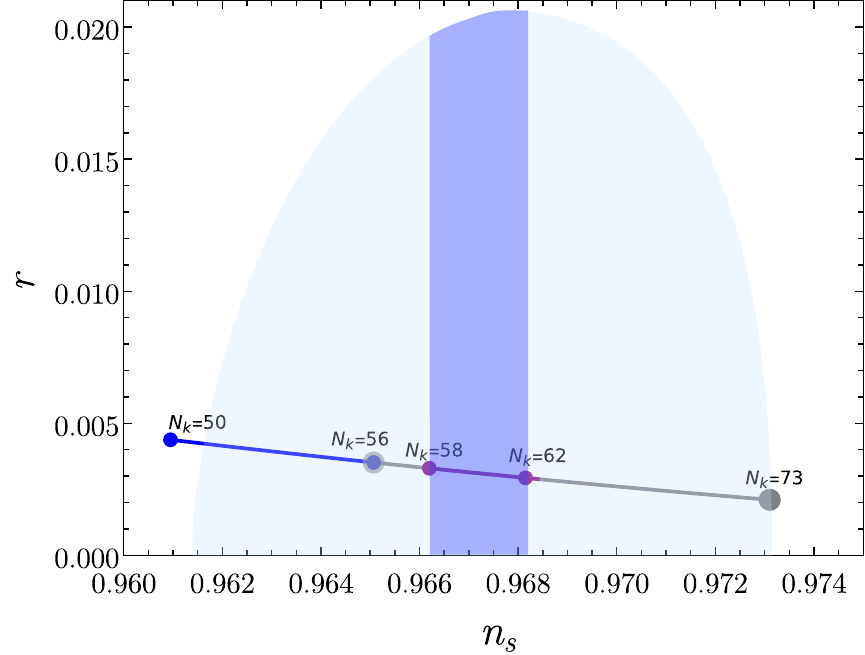}
    \caption{  We plot the inflationary observables, tensor-to-scalar ratio ($r$) as a function of scalar spectral index ($n_s$) for the Starobinsky model using numerical solution given by Eq. \eqref{eq:cmb}. This evolution is presented considering the bound on duration $N_{ k}$ using the reheating regime. We consider different EoS parameters during reheating $w_{\rm re}$ and observe the constraint on the duration $N_{\rm k}$. Correspondingly, in the figure, blue line corresponds to the bound for $w_{\rm re}<1/3$, the gray line for $w_{\rm re}>1/3$ and the purple line for the future observational bound of $n_{\rm s}$. The blue-shaded region represents the $1\sigma$ constraint on the value of $n_{\rm s}$ using ongoing observations \cite{Planck:2018jri, Planck:2018vyg, BICEP:2021xfz, Galloni:2022mok} with $n_{\rm s} = 0.9672 \pm 0.0059$. The dark blue region shows the future projected bound on $n_{\rm s}$ with a sensitivity of $10^{-3}$, assuming its central value remains unchanged.}
    \label{fig: staro nsr ana num leading order}
\end{figure}

We now examine the impact of this on the reheating analysis and compare them with those obtained using slow-roll approximations. This is illustrated in Fig. \ref{fig: staro Nre_Tre ana num leading order}. As anticipated, the improvement is significant. In the figure, one can see that numerical predictions show a significant shift to the right in the estimates of $N_{\rm re}$ and $T_{\rm re}.$ Due to this shift, the bound on $N_k$, and subsequently $n_s$ also changes, with an improvement being $\Delta n_s \sim 10^{-3}$. For example, in the case of leading-order slow-roll approximation with $w_{\rm re} < 1/3,$ the upper bound on $n_s$ is: $n_s \leq 0.964,$ whereas, in this improved case, the bound becomes $ n_s \leq 0.965.$ The improvements are explicitly shown in Table \ref{table:1}. This confirms our anticipation that the accurate dynamics help improve the theoretical predictions, which is one of the main results of this work.

The physical origin of this correction deserves emphasis. In the slow-roll approximation, the end of inflation is defined by $\epsilon_1^{\rm sr} = 1$, where $\epsilon_1^{\rm sr} \equiv (V_{,\phi}/V)^2/2$ is computed purely from the potential. However, as inflation approaches its end, the kinetic energy of the inflaton  $\frac{1}{2}\dot{\phi}^2$ becomes a non-negligible fraction of the total energy density $\rho = \frac{1}{2}\dot{\phi}^2 + V(\phi)$, and the Friedmann equation can no longer be approximated as $3H^2 \simeq V(\phi)$. Consequently, the slow-roll approximation \textit{underestimates} the kinetic contribution near $\epsilon_1 \to 1$, causing the analytical treatment to misidentify the true end of inflation. This leads to a systematic overestimate of $\rho_{\rm end}$ and an underestimate of $N_{\rm end}$, which in turn biases $N_k$  and shifts $n_s$ at the level of $\Delta n_s \sim 10^{-3}$. The full numerical integration of Eqs.~(\ref{eq: H-n}) and (\ref{eq: phi-n}) resolves this by  tracking the exact phase-space trajectory of $(\phi, \dot{\phi})$ without invoking the slow-roll hierarchy, and thus correctly captures the moment 
$\epsilon_1 = 1$ is reached. For completeness we note that other
single-field regimes (for example ultra–slow-roll or constant-roll) alter both
background and perturbation dynamics in different ways; representative studies
include \cite{Shah:2025itu, Morse:2018kda, Odintsov:2019ahz, Dimopoulos:2017ged}. A full treatment of these alternative regimes is beyond the present paper but would be interesting
for future work.

\begin{center}
			\begin{table}[t]
				\begin{tabular}{ |c|c|c|} 
					\hline&&\\
					  & 
					$\hspace{0.5cm} \text{Analytical approximations} \hspace{0.5cm}$&    $\hspace{0.5cm} \text{Numerical solution}\hspace{0.5cm}$  \\[0.8ex] 
					\hline\hline&&\\[-1ex]
					$w_{\rm re} < 1/3$  & $ 52 \leq N_{k} \leq 56 $, &  $ 50 \leq N_{k} \leq 56$,\\[0.3ex]
                         & $ 0.9613 \leq n_{s} \leq 0.9641 $& $ 0.9613 \leq n_{s} \leq 0.9649 $,  \\[0.3ex]
                          &$N_{\rm re} \leq 16$ ($w_{\rm{re}}=0$),&  $N_{\rm re} \leq 21$ ($w_{\rm{re}}=0$),\\[0.3ex]
                        &$T_{\rm re} \geq 10^{10}\ \rm{GeV}$ ($w_{\rm{re}}=0$) &$T_{\rm re} \geq 10^{8}\ \rm{GeV}$ ($w_{\rm{re}}=0$) \\[0.3ex]
					\hline&&\\[-1ex]
					$w_{\rm re} > 1/3$ & $ 56 \leq N_{k} \leq 74 $,&  $ 56 \leq N_{k} \leq 73$,\\[0.3ex]
                        &$ 0.9641 \leq n_{s} \leq 0.9731 $, & $ 0.9649 \leq n_{s} \leq 0.9731 $,   \\[0.3ex]
                        &$N_{\rm re} \leq 37$ ($w_{\rm{re}}=1$),& $N_{\rm re} \leq 34$ ($w_{\rm{re}}=1$),\\[0.3ex]
                        &$T_{\rm re} \geq 10^{-9}$ \rm{GeV} ($w_{\rm{re}}=1$) &$T_{\rm re} \geq 10^{-8}$ \rm{GeV} ($w_{\rm{re}}=1$)  \\[0.3ex]
					\hline&&\\[-1ex]
					future observations & $ 59 \leq N_{k} \leq 63 $,&  $ 58 \leq N_{k} \leq 62$,\\
                       ($w_{\rm re} > 1/3$: allowed) &$ 0.9662 \leq n_{s} \leq 0.9682 $,&$ 0.9662 \leq n_{s} \leq 0.9682 $,\\[0.3ex]
                       ($w_{\rm re}<1/3$: not~~~~~~ &$7 \leq N_{\rm re} \leq 14$ ($w_{\rm{re}}=1$), &$4 \leq N_{\rm re} \leq 12$ ($w_{\rm{re}}=1$),\\[0.3ex]
                        ~~~~~~allowed)&$10^{9}\ {\rm{GeV}} \geq T_{\rm re} \geq 10^{6}\ \rm{GeV}$ ($w_{\rm{re}}=1$) &$10^{12}\ {\rm{GeV}} \geq T_{\rm re} \geq 10^{7}\ \rm{GeV}$ ($w_{\rm{re}}=1$)  \\[0.3ex]
					\hline
				\end{tabular}
				\caption{Starobinsky Inflation: The bounds on variables  $N_{k}, ~n_{s}, ~N_{\rm re},~ \text{and} ~T_{\rm re}$ for different values of $w_{\rm re}$ corresponding to analytically (slow-roll) approximated and the numerical solutions are shown.}
				\label{table:1}
			\end{table}
		\end{center}

\subsection{Implementation of higher-order slow-roll approximations}\label{sec:second_improv}
\begin{figure}[ht]
    \centering
     \includegraphics[width=0.6\textwidth]{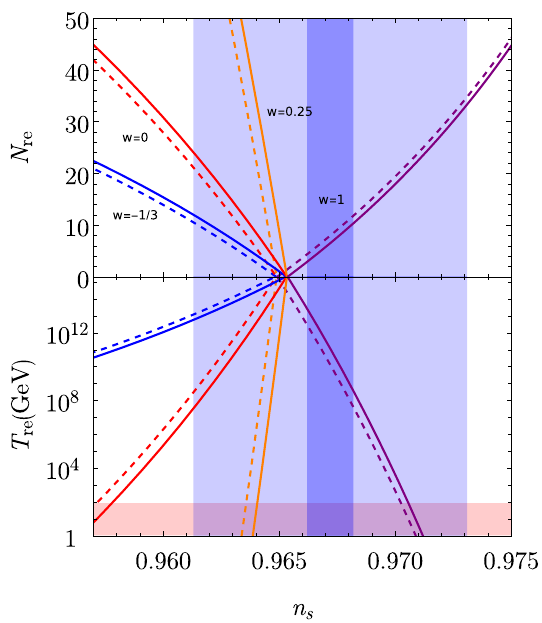}
    \caption{We plot the duration of reheating $N_{\rm re}$ and reheating temperature $T_{\rm re}$ given by Eqs. \eqref{eq: re_duration_final} and \eqref{eq: re temp end} as functions of the scalar spectral index $n_{s}$ parametrically comparing numerical solution with leading-order approximations and numerical solution with higher-order approximations given by Eq. \eqref{eq:cmb}, \eqref{eq: full ns solution} respectively. The solid lines are for the higher-order approximations and the dashed lines are for the leading-order approximations. Please note that different colors represent dynamics corresponding to different effective equations of state parameter $w_{\rm re}$ as indicated in the figure. The blue-shaded region represents the $1\sigma$ constraint on the value of $n_s$ using ongoing observations \cite{Planck:2018jri, Planck:2018vyg, BICEP:2021xfz, Galloni:2022mok} with $n_{s} = 0.9672 \pm 0.0059$. The dark blue region shows the future projected bound on $n_s$  with a sensitivity of $10^{-3}$, assuming its central value remains unchanged. The temperature below the lighter red region is excluded due to the constraint from the electroweak scale, which is taken to be $100$ GeV.
    }
    \label{fig: staro full_app Nre sol}
\end{figure}

\begin{figure}[ht]
    \centering
     \includegraphics[width=0.6\textwidth]{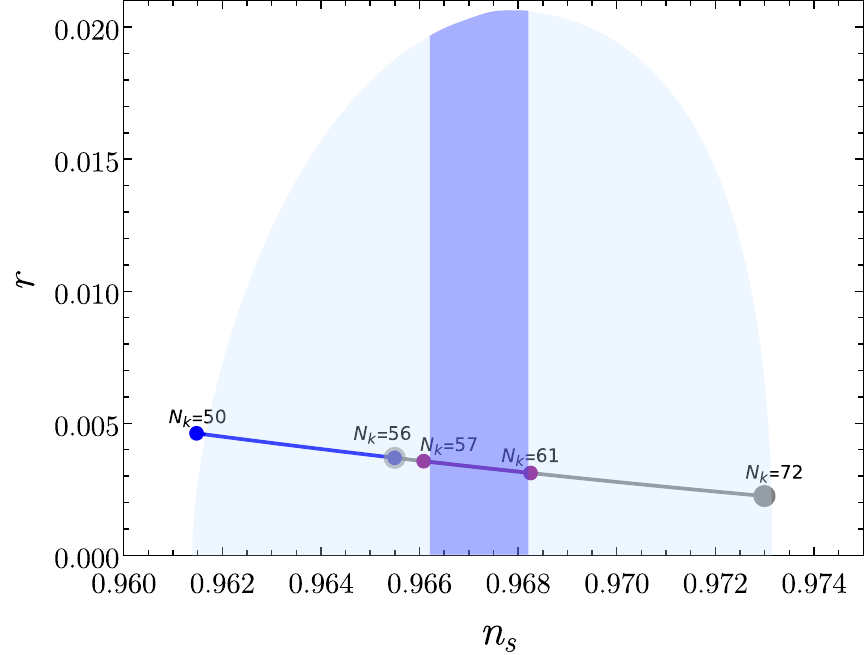}
    \caption{  We plot the inflationary observables, tensor-to-scalar ratio ($r$) as a function of scalar spectral index ($n_s$) for the Starobinsky model using higher-order approximations given by Eq. \eqref{eq: full ns solution} and \eqref{eq: full r solution}. This evolution is presented considering the bound on duration $N_{k}$ using the reheating regime. We consider different EoS parameters during reheating $w_{\rm re}$ and observe the constraint on the duration $N_{ k}$. Correspondingly, in the figure, blue line corresponds to the bound for $w_{\rm re}<1/3$, the gray line for $w_{\rm re}>1/3$ and the purple line for the future observational bound of $n_{\rm s}$. The blue-shaded region represents the $1\sigma$ constraint on the value of $n_{\rm s}$ using ongoing observations \cite{Planck:2018jri, Planck:2018vyg, BICEP:2021xfz, Galloni:2022mok} with $n_{\rm s} = 0.9672 \pm 0.0059$. The dark blue region shows the future projected bound on $n_{\rm s}$ with a sensitivity of $10^{-3}$, assuming its central value remains unchanged.}
    \label{fig: staro full nsr sol}
\end{figure}

In the previous section, we estimated the reheating parameters and e-folding number using the numerical background solution with the leading-order inflationary observables, i.e., Eqs. \eqref{eq:cmb}. However, as mentioned earlier, these expressions are obtained using leading-order slow-roll, and the full solutions of the amplitude of scalar power spectrum, scalar spectral index as well as the tensor-to-scalar ratio in terms of the slow-roll parameter are given as \cite{Huang:2006yt, Martin:2013tda}:
\begin{eqnarray}
    \label{eq: full_As_solution}
     A_{s} &=& 1-2 (C + 1) \epsilon_1 - C \epsilon_2+ \left(2 C^{2} + 2 C + \frac{\pi^2}{2} - f\right) \epsilon^{2}_1 + \left(C^2 - C + \frac{7 \pi^2}{12} - g\right) \epsilon_1 \epsilon_2  \nonumber \\
     &&+ \left(\frac{1}{2} C^2 + \frac{\pi^2}{8}-1 \right) \epsilon^2_{2} + \left(-\frac{1}{2} C^2 + \frac{\pi ^2}{24}\right)\epsilon_2 \epsilon_3, \\
\label{eq: full ns solution}
 n_{s} &=& 1 - 2 \epsilon_1 - \epsilon_2 -2 \epsilon^2_1 - (3 + 2C)\epsilon_1 \epsilon_2 - C\epsilon_2 \epsilon_3, \\
 \label{eq: full r solution}
    r &=& 16 \epsilon_1 (1 + 2 C \frac {\epsilon_2}{2}),
\end{eqnarray}
where $C = -2 + \ln{2} +\gamma$, and $\gamma$ is the Euler constant, $f=5$ and $g=7$. Here, all slow-roll parameters are evaluated using the full numerical background solution. In this section, instead of using Eqs. \eqref{eq:cmb}, along with the numerical solution, we consider the above expressions for the observable, repeat the reheating analysis, and compare our results with the numerical results obtained in the previous section, which has been illustrated in Figs. \ref{fig: staro full_app Nre sol} and \ref{fig: staro full nsr sol} as well as in Table \ref{table:2}.

As can be seen again in Fig. \ref{fig: staro full_app Nre sol}, there is a further shift in reheating parameters, $N_{\rm re}$ and $T_{\rm re}$ compared to the improvements mentioned in the previous section. We obtain an additional  improvement of $ \Delta n_s \sim 4 \times 10^{-4},$ resulting in a new bound on $n_s$, namely $n_s \leq 0.9653$ for $w_{\rm re} < 1/3.$ Thus, considering higher-order slow-roll approximations instead of the leading-order leads to an additional correction of order $4 \times 10^{-4}$, further refining the theoretical predictions. The detailed bounds on parameters are shown in the Table \ref{table:2}.

To understand this physically, recall that the scalar spectral index is evaluated 
at the moment the pivot scale $k_\ast = 0.05~\text{Mpc}^{-1}$ crosses the Hubble 
horizon during inflation, i.e. at $k_\ast = a_\ast H_\ast$. At this moment, 
the slow-roll parameters $\epsilon_1$ and $\epsilon_2$ are already non-zero --- 
for the Starobinsky model with $N_k \sim 55$, one finds $\epsilon_1 \sim 10^{-3}$ 
and $\epsilon_2 \sim 2/N_k \sim 0.036$. Truncating the perturbation spectrum at 
leading-order ($n_s \simeq 1 - 2\epsilon_1 - \epsilon_2$) therefore neglects 
contributions of order $\epsilon_1^2$, $\epsilon_1\epsilon_2$, $\epsilon_2^2$, 
and $\epsilon_3$, which are not identically zero at horizon crossing. The 
higher-order corrections~\cite{Stewart:1993bc} systematically account 
for these terms, tightening the prediction for $n_s$. The resulting shift of 
$\sim 4 \times 10^{-4}$ is subdominant compared to the numerical background 
correction ($\sim 10^{-3}$) but is nonetheless at the threshold of detectability 
for next-generation CMB experiments.

\begin{center}
			\begin{table}[t]
				\begin{tabular}{ |c|c|c|} 
					\hline&&\\
					&$\hspace{1.0cm}$  Numerical solution with $\hspace{1.0cm}$&$\hspace{1.0cm}$ Numerical solution with $\hspace{1.0cm}$ \\
                        &  leading-order $n_s$, $r$    &   higher-order $n_s$, $r$ \\ & expressions & expressions \\[0.8ex] 
					\hline\hline&&\\[-1ex]
					$w_{\rm re} < 1/3$ & $ 50 \leq N_{k} \leq 56 $, &  $ 50 \leq N_{k} \leq 56$,\\[0.3ex]
                         & $ 0.9613 \leq n_{s} \leq 0.9649 $& $ 0.9613 \leq n_{s} \leq 0.9653 $,  \\[0.3ex]
                          &$N_{\rm re} \leq 21$ $(w_{\rm re}=0)$,&  $N_{\rm re} \leq 24$$(w_{\rm re}=0)$,\\[0.3ex]
                        &$T_{\rm re} \geq 10^{8}\ \rm{GeV}$$(w_{\rm re}=0)$ &$T_{\rm re} \geq 10^{7}\ \rm{GeV}$ $(w_{\rm re}=0)$ \\[0.3ex]
					\hline&&\\[-1ex]
					$w_{\rm re} > 1/3$ & $ 56 \leq N_{k} \leq 73 $,&  $ 56 \leq N_{k} \leq 72$,\\[0.3ex]
                        &$ 0.9649 \leq n_{s} \leq 0.9731 $, & $ 0.9653 \leq n_{s} \leq 0.9731 $,   \\[0.3ex]
                        &$N_{\rm re} \leq 34$$(w_{\rm re}=1)$,& $N_{\rm re} \leq 33$$(w_{\rm re}=1)$,\\[0.3ex]
                        &$T_{\rm re} \geq 10^{-8}$ \rm{GeV}  $(w_{\rm re}=1)$&$T_{\rm re} \geq 10^{-7}\ \rm{GeV}$ $(w_{\rm re}=1)$ \\[0.3ex]
					\hline&&\\[-1ex]
					future observations & $ 58 \leq N_{k} \leq 62 $,&  $ 57 \leq N_{k} \leq 61$,\\
                       ($w_{\rm re} > 1/3$: allowed) &$ 0.9662 \leq n_{s} \leq 0.9682 $,&$ 0.9662 \leq n_{s} \leq 0.9682 $,\\[0.3ex]
                        ($w_{\rm re} < 1/3$: not ~~~~~~&$4 \leq N_{\rm re} \leq 12$ $(w_{\rm re}=1)$, &$3 \leq N_{\rm re} \leq 10$ $(w_{\rm re}=1)$,\\[0.3ex]
                       ~~~~~~ allowed)&$10^{12}\ \rm{GeV} \geq T_{\rm re} \geq 10^{7}\ \rm{GeV}$ $(w_{\rm re}=1)$&$10^{13}\ \rm{GeV} \geq T_{\rm re} \geq 10^{8}\ \rm{GeV}$ $(w_{\rm re}=1)$ \\[0.3ex]
				
					\hline
				\end{tabular}
				\caption{Starobinsky Inflation: The bounds on variables  $N_{k}, ~n_{s}, ~N_{\rm re},~\text{and}  \ T_{\rm re}$ for different values of $w_{\rm re}$ corresponding to leading-order approximations and the higher-order approximations are shown. Note that, in both cases, the background solutions are numerically obtained.}
				\label{table:2}
			\end{table}
		\end{center}


\subsection{Implementation of the onset of reheating as the bottom of the potential}\label{sec:third_improv}

Thus far, we were estimating $N_{\rm re}$ and $T_{\rm re}$, considering the onset of the reheating era after the end of inflation. However, in Ref. \cite{Nandi:2019xve}, it has been shown that, instead of considering minimal gravity theory, if one considers modified gravity theories, the onset of reheating as the end of inflation can make a significant change in the analysis. This is because, $\epsilon_1$ is not an invariant under conformal transformation. Therefore, in a meaningful manner, one can choose the onset of reheating as the bottom of the potential, as this point is conformally invariant, and can be used even for general modified gravity theories. Therefore, in this section, we introduce corrections to the reheating analysis, and thus the observables, by considering that the reheating era commences after the epoch at which the potential touches its minima (bottom of the potential). In that case, Eq. \eqref{eq: re_duration_final}  can be re-written as
\begin{eqnarray}
\label{eq: re duration bottom}
 N_{\rm re} &=& \frac{4}{1 - 3 w_{\text {re}}} \left[-N_k - N_{\text {eb}} - \frac{1}{4} \ln \left(\frac{30}{\pi^2 g_{\text {reh}}}\right) - \frac{1}{3} \ln \left(\frac{11 g_{\text {s,re}}}{43}\right)- \ln \left(\frac{k}{a_0 T_0}\right) \right.\nonumber\\
 && \left. - \frac{1}{4} \ln \left(\frac{\rho_{\text {b}}}{H_k^4}\right)\right],
\end{eqnarray}
and the reheating temperature, i.e., Eq. \eqref{eq: re temp end} can also be evaluated as:
\begin{eqnarray}
\label{eq: re temp bottom}
T_{\text {re}} = \frac{a_0 T_0}{k}\left(\frac{43}{11 g_{\text {s,re}}}\right) e^{-N_{\text k} - N_{\text {eb}} - N_{\text {re}}}.
\end{eqnarray}

\noindent Here, $N_{\rm eb} \equiv \ln (a_{\rm b}/a_{\rm end})$ and $\rho_{\rm b} \equiv 3 H_{\rm b}^2,$ where $(b)$ denotes the bottom of the potential. Again, we perform our analysis considering the numerically obtained solution with higher-order slow-roll corrections and the onset of reheating as the bottom of the potential. However, to our surprise, we find that this implementation does not improve the accuracy to a significant amount, as $\Delta n_s \sim 10^{-5}.$ This small correction reflects the fact that, in minimally coupled models, the difference between the end of inflation and the minimum of the potential is dynamically short.

This small contribution can be understood as follows. The reheating onset 
	determines the initial value of $\rho_{\rm re,i}$ used in the computation of 
	$N_{\rm re}$ via Eq.~(\ref{eq: re_duration_final}). When the onset is shifted from $\phi_{\rm end}$ 
	(where $\epsilon_1 = 1$) to $\phi_b$ (where $V'(\phi_b) = 0$, the potential 
	minimum), the inflaton has already transferred a fraction of its energy to 
	oscillations. However, for the Starobinsky potential, the ratio 
	$\rho_{\rm end}/\rho_b \simeq 1 + \mathcal{O}(\epsilon_1^2)$ at the transition 
	point, meaning the energy difference between these two definitions is at most 
	second order in slow-roll. Since $N_{re} \propto \ln(\rho_{\rm re,i}/\rho_{\rm re,f})$, 
	the logarithmic sensitivity of $N_{\rm re}$ to the initial condition suppresses the 
	shift further, resulting in a negligible $\Delta n_s \sim 10^{-5}$. This confirms 
	that while careful treatment of the reheating \textit{onset} is conceptually 
	important, the dominant source of theoretical uncertainty lies in the accurate 
	determination of $\rho_{\rm end}$ through the full numerical background evolution.

\begin{figure}[h!]
    \centering
      \includegraphics[width=0.6\textwidth]{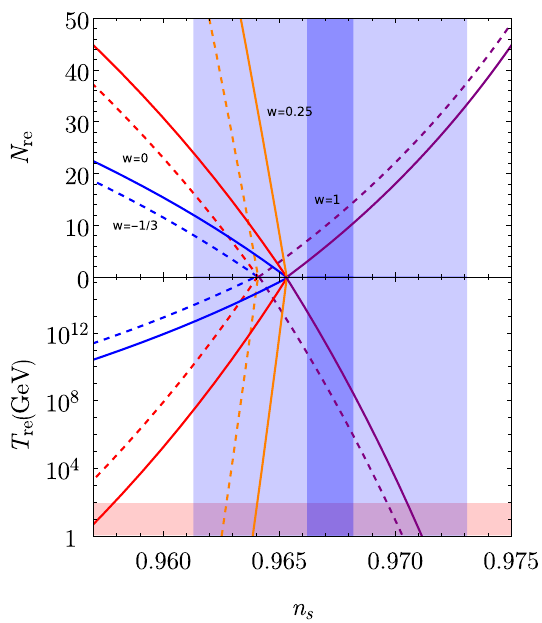}
    \caption{We plot the duration of reheating $N_{\rm re}$ and reheating temperature $T_{\rm re}$ given by Eqs. \eqref{eq: re_duration_final} and \eqref{eq: re temp end} as functions of the scalar spectral index $n_{s}$ parametrically comparing analytically approximated solution and numerical background solution with higher-order slow-roll corrections, given by Eqs. \eqref{eq: starobinsky observable parameters}, \eqref{eq: full ns solution} respectively. The solid lines are for the onset of reheating as the bottom of the potential and the dashed lines are for the analytically approximated solution. Please note that different colors represent dynamics corresponding to different effective equations of state parameter $w_{\rm re}$ as indicated in the figure. The blue-shaded region represents the $1\sigma$ constraint on the value of $n_s$ using ongoing observations \cite{Planck:2018jri, Planck:2018vyg, BICEP:2021xfz, Galloni:2022mok} with $n_{s} = 0.9672 \pm 0.0059$. The dark blue region shows the future projected bound on $n_s$  with a sensitivity of $10^{-3}$, assuming its central value remains unchanged. The temperature below the lighter red region is excluded due to the constraint from the electroweak scale, which is taken to be $100$ GeV.
    }
    \label{fig: staro bottom Nre_Tre}
\end{figure}


To summarize our cumulative result, in Fig. \ref{fig: staro bottom Nre_Tre}, we plot the results obtained through the analytical approximations mentioned in the first section and the improvement obtained through the implementation of three corrections to the dynamics, and Table \ref{table:3} summarizes the cumulative effect of all corrections relative to the purely analytical treatment. Compared to the purely analytical slow-roll treatment presented in Sec. \ref{sec: reheating}, the cumulative implementation of these corrections results in a maximum shift of $\Delta n_s \sim 1.2 \times 10^{-3}$ within the allowed reheating range. This shift is evaluated relative to the leading-order slow-roll prediction at fixed reheating assumptions. Note that, if the central value of $n_s$ is assumed to be $\sim 0.9672$ even in this case, then even with the improvements, the Starobinsky model would be disfavored at the $1 \sigma$ level from the expected future observation. Although illustrated using the Starobinsky model, similar end-of-inflation corrections are expected in any inflationary model where slow-roll breaks down near the end of inflation. 

It is instructive to compare the present results quantitatively with prior treatments 
	of the Starobinsky model in the literature. Within the standard analytical slow-roll 
	approximation $n_s \simeq 1 - 2/N_k$, the leading-order treatments of 
	Cook et al.~\cite{Cook:2015vqa} and Martin \& Ringeval~\cite{Martin:2010kz} yield 
	$n_s \leq 0.9641$ for $w_{re} < 1/3$, corresponding to $N_k \leq 56$, and 
	$N_{re} \leq 16$, $T_{re} \geq 10^{10}$~GeV for $w_{re} = 0$, as reproduced in 
	our Table~\ref{table:3} (left column). Our improved framework --- incorporating 
	full numerical background evolution, higher-order slow-roll corrections, and a revised 
	onset of reheating at the potential minimum --- shifts this upper bound to 
	$n_s \leq 0.9653$, corresponding to $N_{re} \leq 24$ and $T_{re} \geq 10^{7}$~GeV 
	for $w_{re} = 0$ (Table~\ref{table:3}, right column), a cumulative upward shift 
	of $\Delta n_s \sim 1.2 \times 10^{-3}$. A related refinement in the same direction 
	was recently reported by Drees \& Xu~\cite{Drees:2025ngb}, who derive an improved 
	analytical approximation
	\begin{equation}
		n_s \simeq 1 - \frac{2}{N_\star - \dfrac{3}{4}\ln\!\left(\dfrac{2}{N_\star}
			\right)},
	\end{equation}
and demonstrate that the standard expression $n_s \simeq 1 - 2/N_k$ systematically underestimates $n_s$ for the Starobinsky model, consistent with the direction of our corrections. However, unlike that work, the present analysis employs a full numerical integration of the background equations~(Eqs.~(\ref{eq: H-n}) and (\ref{eq: phi-n})) and additionally decomposes each correction individually: the numerical background treatment alone contributes $\Delta n_s \sim 10^{-3}$, higher-order slow-roll corrections add a further $\sim 4 \times 10^{-4}$, and the revised onset of reheating contributes negligibly at $\sim 10^{-5}$. For $w_{re} > 1/3$, the upper bound on $N_k$ shifts from $N_k \leq 74$ (analytical) to $N_k \leq 72$ (corrected), while the corresponding lower bound on $n_s$ shifts from $0.9641$ to $0.9653$. The correction $\Delta n_s \sim 10^{-3}$ is comparable in magnitude to the projected sensitivity of forthcoming experiments such as PRISM~\cite{Andre:2013afa}, EUCLID~\cite{Amendola:2012ys}, CORE~\cite{Finelli:2016cyd}, and cosmic 21-cm surveys~\cite{Mao:2008ug}, underscoring that theoretical precision at this level is indispensable for unambiguous model discrimination in the next generation of CMB observations. To our knowledge, a systematic decomposition of end-of-inflation corrections and their explicit propagation into $n_s$ has not been reported in prior reheating analyses of the Starobinsky model. These are the main results of this work. 

This demonstrates that theoretical control at the $\mathcal{O}(10^{-3})$ level is necessary for robust model discrimination: models marginally accepted or excluded under leading-order slow-roll predictions may move across acceptance boundaries once end-of-inflation corrections are included. A dedicated forecast that folds our $\Delta n_s$ corrections into Fisher- or MCMC-based model-selection would be a valuable follow-up study.

 \begin{center}
     
			\begin{table}[h]
				\begin{tabular}{ |c|c|c|} 
					\hline&&\\
					  & 
					$\hspace{1.0cm}\text{Analytical} \hspace{1.0cm}$&$\hspace{1.0cm} \text{Higher-order}\hspace{1.0cm}$ \\
                         & 
					approximations&      approximations  \\[0.8ex] 
					\hline\hline&&\\[-1ex]
					$w_{\rm re} < 1/3$  & $ 52 \leq N_{k} \leq 56 $, &  $ 50 \leq N_{k} \leq 56$,\\[0.3ex]
                         & $ 0.9613 \leq n_{s} \leq 0.9641 $,& $ 0.9613 \leq n_{s} \leq 0.9653 $,  \\[0.3ex]
                          &$N_{\rm re} \leq 16$ $(w_{\rm re}=0)$,&  $N_{\rm re} \leq 24$ $(w_{\rm re}=0)$,\\[0.3ex]
                        &$T_{\rm re} \geq 10^{10}\ \rm{GeV}$ $(w_{\rm re}=0)$&$T_{\rm re} \geq 10^{7}\ \rm{GeV}$ $(w_{\rm re}=0)$ \\[0.3ex]
					\hline&&\\[-1ex]
					$w_{\rm re} > 1/3$ & $ 56 \leq N_{k} \leq 74 $,&  $ 56 \leq N_{k} \leq 72$,\\[0.3ex]
                        &$ 0.9641 \leq n_{s} \leq 0.9731 $, & $ 0.9653 \leq n_{s} \leq 0.9731 $,   \\[0.3ex]
                       &$N_{\rm re} \leq 37$ $(w_{\rm re}=1)$,&  $N_{\rm re} \leq 33$ $(w_{\rm re}=1)$,\\[0.3ex]
                        &$T_{\rm re} \geq 10^{-9}\ \rm{GeV}$ $(w_{\rm re}=1)$&$T_{\rm re} \geq 10^{-7}\ \rm{GeV}$ $(w_{\rm re}=1)$  \\[0.3ex]
					\hline&&\\[-1ex]
					future observations & $ 59 \leq N_{k} \leq 63 $,&  $ 57 \leq N_{k} \leq 61$,\\
                        ($w_{\rm re} > 1/3$: allowed) &$ 0.9662 \leq n_{s} \leq 0.9682 $,&$ 0.9662 \leq n_{s} \leq 0.9682 $,\\[0.3ex]
                       ($w_{\rm re} < 1/3$: not~~~~~~&$7 \leq N_{\rm re} \leq 14$ $(w_{\rm re}=1)$, &$3 \leq N_{\rm re} \leq 10$ $(w_{\rm re}=1)$,\\[0.3ex]
                        ~~~~~~allowed)&$10^{9}\ \rm{GeV} \geq T_{\rm re} \geq 10^{6}\ \rm{GeV}$  $(w_{\rm re}=1)$&$10^{13}\ \rm{GeV} \geq T_{\rm re} \geq 10^{8}\ \rm{GeV}$  $(w_{\rm re}=1)$\\[0.3ex]
					\hline
				\end{tabular}
				\caption{Starobinsky Inflation: The bounds on variables,  $N_{k}, ~n_{s}, ~N_{\rm re},~ \text{and} ~T_{\rm re}$ for different values of $w_{\rm re}$ corresponding to analytical approximated parameters, and numerical solution with higher-order approximations are shown.}
				\label{table:3}
			\end{table}
		\end{center}

\section{Summary and conclusions}\label{sec:conclu}
In this article, we consider a single canonical scalar field model minimally coupled to the gravity with a potential $V(\phi)$ that drives the evolution of the early universe, including both slow-roll inflation and oscillatory solution around the minimum of the potential, also known as the reheating epoch. In obtaining the solution, we often use different sets of approximations in both regimes. In the slow-roll inflationary regime, where $\epsilon_1, \epsilon_2 \ll 1,$ we rely upon the slow-roll approximations. In contrast, the reheating epoch is, for simplicity, quantitatively characterized by the effective EoS parameter $w_{\rm re}$ and the duration of this epoch $N_{\rm re}$ (or the reheating temperature $T_{\rm re}$). However, these approximations may lead to a significant discrepancy in the theoretical estimation of the observables ($n_s$ and $r$). While accurately defining the reheating epoch (\viz the qualitative analysis) is challenging, this work focuses on examining the impact of a more accurate dynamics, rather than the slow-roll approximations, on the perturbed observables.

In order to do this, we specifically considered the following improvements in the dynamics:

\begin{enumerate}
    \item Numerical Solution: In this method, we solve the complete background Eq. \eqref{eq: phi-n} using the numerical method, which provides us with an accurate solution for the dynamics of the universe in its early phase. Using this solution, we estimate the observables and compare the results with the solutions obtained using slow-roll approximation.
    \item Higher-order approximations: This method includes higher-order slow-roll corrections to the inflationary observables, Eqs. \eqref{eq: full_As_solution}, \eqref{eq: full ns solution} and \eqref{eq: full r solution}, rather than Eqs. \eqref{eq:cmb}. Here also, we used complete numerical solution and estimated the observables using higher-order slow-roll corrections and estimated the discrepancy in the observables with respect to the analytical solution obtained using slow-roll approximation.
    \item Onset of the reheating as the bottom of the potential: Typically, the end of inflation is considered the onset of the reheating epoch. However, in this method, we show that if reheating begins at the bottom of the potential, the reheating parameters are modified, as given by Eq. \eqref{eq: re duration bottom} and Eq. \eqref{eq: re temp bottom}. This modification can provide a new bound on the duration $N_k$, leading to a discrepancy in the scalar spectral index.
\end{enumerate}

 After implementing these changes, we found an improvement in the theoretical bounds on the observables. As seen in Fig. \ref{fig: staro bottom Nre_Tre}, there is an improvement of $\Delta n_s \sim 1.2 \times 10^{-3}.$  These numerical predictions indicate that for $w_{\rm re}<1/3$, the Starobinsky model has an upper bound as $n_s \leq 0.9653.$ This can be disfavored from the future observational constraint if the central value is $n_s \simeq 0.9672$ \cite{Galloni:2022mok}. However, for $w_{\rm re}>1/3$, the reheating parameters fall within the future observational bounds, though, at the cost of sufficient reheating e-folding number ($\sim \mathcal{O}(1 - 2)$). This scenario would be strongly constrained if future observations favor instantaneous reheating \cite{Felder:1998vq, deHaro:2023xcc}.
 
 Physically, these corrections share a common origin: the standard slow-roll approximation is inherently perturbative and assumes that $\epsilon_i \ll 1$ throughout inflation. However, all three corrections become relevant precisely in the regime where this assumption weakens --- near the end of inflation where $\epsilon_1 \to 1$ (numerical correction), at horizon crossing where $\epsilon_2 \not\ll 1$ (higher-order correction), and at the transition to the oscillatory phase (onset correction). The Starobinsky model, with its characteristic plateau potential, is particularly sensitive to these effects because the slow-roll parameters evolve rapidly in the final $\mathcal{O}(1)$ e-fold before inflation ends. Models with similarly steep exit from the inflationary plateau --- such as $\alpha$-attractors and Higgs inflation --- are expected to exhibit corrections of comparable magnitude, making the present analysis broadly relevant beyond the specific case studied here. By contrast, models whose exit dynamics are much smoother or whose reheating is driven by qualitatively different microphysics (small-field models with very flat exits, or scenarios with early preheating) may show smaller or model-dependent shifts. A practical conclusion is that the numerical re-evaluation we propose is straightforward to apply model-by-model, and should be performed when aiming for theoretical predictions at $\mathcal{O}(10^{-3})$ precision.

These results demonstrate that an accurate theoretical treatment of end-of-inflation dynamics is as essential as higher-order perturbative corrections in precision-era inflationary analyses. Even corrections that are often regarded as subleading can induce observable shifts in perturbed quantities, and therefore must be systematically incorporated when deriving constraints on inflationary models. While the present work focuses on slow-roll dynamics within a minimal framework, further refinements --- including a more detailed reheating treatment, higher-order correlations such as non-Gaussianity \cite{Maldacena:2002vr, Nandi:2015ogk, Nandi:2016pfr}, the role of primordial black holes and gravitational waves, extensions to non-minimal gravity \cite{Nandi:2019xlj}, and alternative early-universe scenarios such as viable classical bounces \cite{Nandi:2019xag, Nandi:2020sif, Nandi:2020szp, Nandi:2022twa, Nandi:2023ooo, Kaur:2023uaz} --- may further sharpen theoretical predictions. It would also be valuable to extend our numerical framework to warm inflation (introducing a dissipation term $\Gamma(\phi, T)$ and a coupled radiation component) and quantify whether dissipative dynamics amplify or suppress the $\Delta n_s \sim 10^{-3}$ correction found here for cold inflation. Such a direct warm-cold comparison will show whether both paradigms produce comparable-order shifts in observables or whether dissipation qualitatively changes the sensitivity to end-of-inflation modelling. A careful and consistent implementation of such effects remains an important direction for future investigations.

\section*{Acknowledgements}
DN is supported by the DST, Government of India through the DST-INSPIRE Faculty fellowship (04/2020/002142). MK is supported by a DST-INSPIRE Fellowship under the reference number: IF170808, DST, Government of India. D.N. is grateful to the Department of Physics, School of Advanced Sciences, Vellore Institute of Technology (VIT) Chennai, for institutional support and academic encouragement. DN, SY and MK are also very thankful to the Department of Physics and Astrophysics, University of Delhi. MK, SY and DN also acknowledge facilities provided by the IUCAA Centre for Astronomy Research and Development (ICARD), University of Delhi.

\end{document}